\documentclass[prl,superscriptaddress,amsmath,amssymb]{revtex4}
\usepackage{color,graphicx}
\usepackage{epsfig,amsmath}
\usepackage[utf8]{inputenc}
               
\begin{document}
 
\title{Production of $\alpha$-particle 
condensate states in heavy-ion collisions}

\author{Ad. R. Raduta}
\affiliation{Institut de Physique Nucl\'eaire, CNRS/IN2P3,
Universit\'e Paris-Sud 11, Orsay, France}
\affiliation{National Institute for Physics and Nuclear Engineering,
Bucharest-Magurele, Romania}
\author{B.~Borderie}
\affiliation{Institut de Physique Nucl\'eaire, CNRS/IN2P3,
Universit\'e Paris-Sud 11, Orsay, France}
\author{E. Geraci}
\affiliation{INFN, Sezione di Catania, Italy}
\affiliation{Dipartimento di Fisica e Astronomia,
Universit\`a di Catania, Italy}
\affiliation{INFN, Sezione di Bologna and Dipartimento di Fisica,
Universit\`a di Bologna, Italy}
\author{N. Le Neindre}
\affiliation{Institut de Physique Nucl\'eaire, CNRS/IN2P3,
Universit\'e Paris-Sud 11, Orsay, France}
\affiliation{LPC, CNRS/IN2P3, Ensicaen, Universit\'{e} de Caen, 
Caen, France}
\author{P.~Napolitani}
\affiliation{Institut de Physique Nucl\'eaire, CNRS/IN2P3,
Universit\'e Paris-Sud 11, Orsay, France}
\author{M.~F.~Rivet}
\affiliation{Institut de Physique Nucl\'eaire, CNRS/IN2P3,
Universit\'e Paris-Sud 11, Orsay, France}
\author{R.~Alba}
\affiliation{INFN, Laboratori Nazionali del Sud, Italy}
\author{F. Amorini}
\affiliation{INFN, Laboratori Nazionali del Sud, Italy}
\author{G. Cardella}
\affiliation{INFN, Sezione di Catania, Italy}
\author{M. Chatterjee}
\affiliation{Saha Institute of Nuclear Physics,
Kolkata, India}
\author{E. De Filippo}
\affiliation{INFN, Sezione di Catania, Italy}
\author{D.~Guinet}
\affiliation{Institut de Physique Nucl\'eaire, CNRS/IN2P3,
Universit\'e Claude Bernard Lyon 1, Villeurbanne, France}
\author{P.~Lautesse}
\affiliation{Institut de Physique Nucl\'eaire, CNRS/IN2P3,
Universit\'e Claude Bernard Lyon 1, Villeurbanne, France}
\author{E. La Guidara}
\affiliation{INFN, Sezione di Catania, Italy}
\affiliation{CSFNSM, Catania, Italy}
\author{G. Lanzalone}
\affiliation{INFN, Laboratori Nazionali del Sud, Italy}
\affiliation{Universit\`a di Enna ``Kore'', Enna, Italy}
\author{G. Lanzano}
\altaffiliation{deceased}
\affiliation{INFN, Sezione di Catania, Italy}
\author{I. Lombardo}
\affiliation{INFN, Laboratori Nazionali del Sud, Italy}
\affiliation{Dipartimento di
Fisica e Astronomia, Universit\`a di Catania, Italy}
\author{O.~Lopez}
\affiliation{LPC, CNRS/IN2P3, Ensicaen, Universit\'{e} de Caen, 
Caen, France}
\author{C.~Maiolino}
\affiliation{INFN, Laboratori Nazionali del Sud, Italy}
\author{A.~Pagano}
\affiliation{INFN, Sezione di Catania, Italy}
\author{S. Pirrone}
\affiliation{INFN, Sezione di Catania, Italy}
\author{G. Politi}
\affiliation{INFN, Sezione di Catania, Italy}
\affiliation{Dipartimento di Fisica e Astronomia,
Universit\`a di Catania, Italy}
\author{F. Porto}
\affiliation{INFN, Laboratori Nazionali del Sud, Italy}
\affiliation{Dipartimento di
Fisica e Astronomia, Universit\`a di Catania, Italy}
\author{F. Rizzo}
\affiliation{INFN, Laboratori Nazionali del Sud, Italy}
\affiliation{Dipartimento di
Fisica e Astronomia, Universit\`a di Catania, Italy}
\author{P. Russotto}
\affiliation{INFN, Laboratori Nazionali del Sud, Italy}
\affiliation{Dipartimento di
Fisica e Astronomia, Universit\`a di Catania, Italy}
\author{J.P.~Wieleczko}
\affiliation{GANIL, (DSM-CEA/CNRS/IN2P3), Caen, France}

\begin{abstract}
The fragmentation of quasi-projectiles from the nuclear reaction
$^{40}Ca$+$^{12}C$ at 25 MeV/nucleon was used to produce excited states
candidates to $\alpha$-particle condensation. 
The experiment was performed at LNS-Catania using the CHIMERA 
multidetector.
Accepting the emission simultaneity and 
equality among the $\alpha$-particle kinetic energies as experimental criteria
for deciding in favor of the condensate nature of an excited state,
we analyze the $0_2^+$ and $2_2^+$ states of $^{12}$C and the 
$0_6^+$ state of  $^{16}$O.
A sub-class of events corresponding to the direct 3-$\alpha$ decay 
of the Hoyle state is isolated.
\end{abstract}

\maketitle


Alpha particle condensed states have been predicted 
to exist in self-conjugate 4N nuclei
in the vicinity of the $N\alpha$-decay threshold and are characterized by
a low density \cite{Tohsaki_PRL2001,Suzuki_PRC2002,Yamada_PRC2004}.
Theoretical predictions agree that the most probable candidate states  
correspond to the Hoyle state
and $2_2^+$ state of $^{12}$C \cite{Funaki_EPJA2005,Yamada_EPJA2005}
and to the $0_6^+$ state of $^{16}O$ \cite{Funaki_PRL2008}.
If confirmed, these states should be regarded as the 
finite size replica of 4-nucleon clusterization taking place in 
dilute symmetric nuclear matter \cite{Ropke_PRL1998,Beyer_PLB2000}.

From the experimental point of view,
a candidate for $\alpha$-particle condensation
is a resonant state whose excitation energy is close to the
$N\alpha$ threshold and which directly decays into $N\alpha$
of equal kinetic energies (in the center of mass of the emitter). 
Once agreed on these criteria, it comes out that, 
probably, the most appropriate working methodology involves full 
4$\pi$ detection of energetic reaction products.
High angular granularity is an extra requirement for having
high kinematic resolution and detection efficiency.

Data discussed in this paper have been collected in the nuclear reaction 
$^{40}$Ca+$^{12}$C at 25 MeV per nucleon incident energy and correspond to 
quasi-projectile (QP) decay events.
The experiment was performed at LNS-Catania (Italy),
using the CHIMERA \cite{chimera} multidetector. 
Details of the experimental settings and methods for reaction product
identification and calibration may be found in Ref. \cite{prl}

The overall quality of the energy calibration was judged 
based on the energy evaluation of several resonant
states of light nuclei inferred from two particle correlation functions
($\alpha-\alpha$, $\alpha-d$, $\alpha-t$) \cite{prl}.
Taking into account the complexity of the apparatus and the fact that
the maximum obtained deviation was of 20 keV, 
the calibration was considered reliable enough
to allow for the due spectroscopic analyses.


Correlation functions are known to reveal also space-time information on the
decaying source taking advantage of proximity effects induced by Coulomb
repulsion and to highlight the production of event classes with specific
properties. Their predictive power is now exploited in order to qualify 
the simultaneity/sequentiality character of a specific state decay.
Let us define a generalized correlation function in terms of $\alpha$-particle
average kinetic energy and RMS,
\begin{equation} 
1+R(\left< E_{\alpha}  \right>,\sigma_{E_{\alpha}})=
\frac{Y_{corr}(\left< E_{\alpha}
  \right>,\sigma_{E_{\alpha}})}{Y_{uncorr}(\left< E_{\alpha}
  \right>,\sigma_{E_{\alpha}})},
\label{eq:intraev}
\end{equation}
where $Y_{corr}$ and $Y_{uncorr}$ represent respectively the correlated and
uncorrelated number of events corresponding to given values of
$(\left< E_{\alpha}\right>,\sigma_{E_{\alpha}})$,
and see how does it look like for 3-$\alpha$ decay  events.

\begin{figure}[th]
\centerline{
\vspace*{-10pt}
\psfig{file=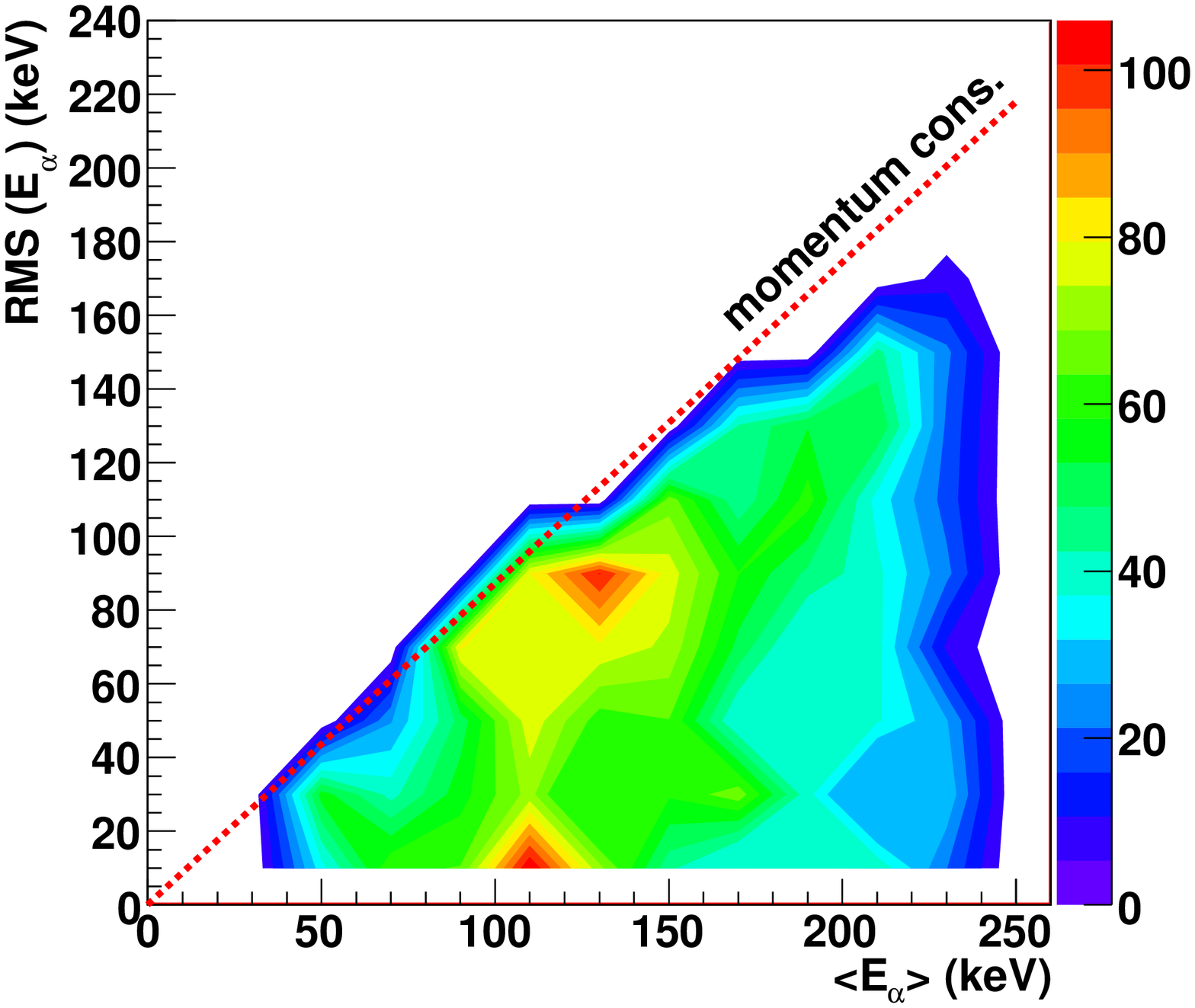,width=7.8cm}
\psfig{file=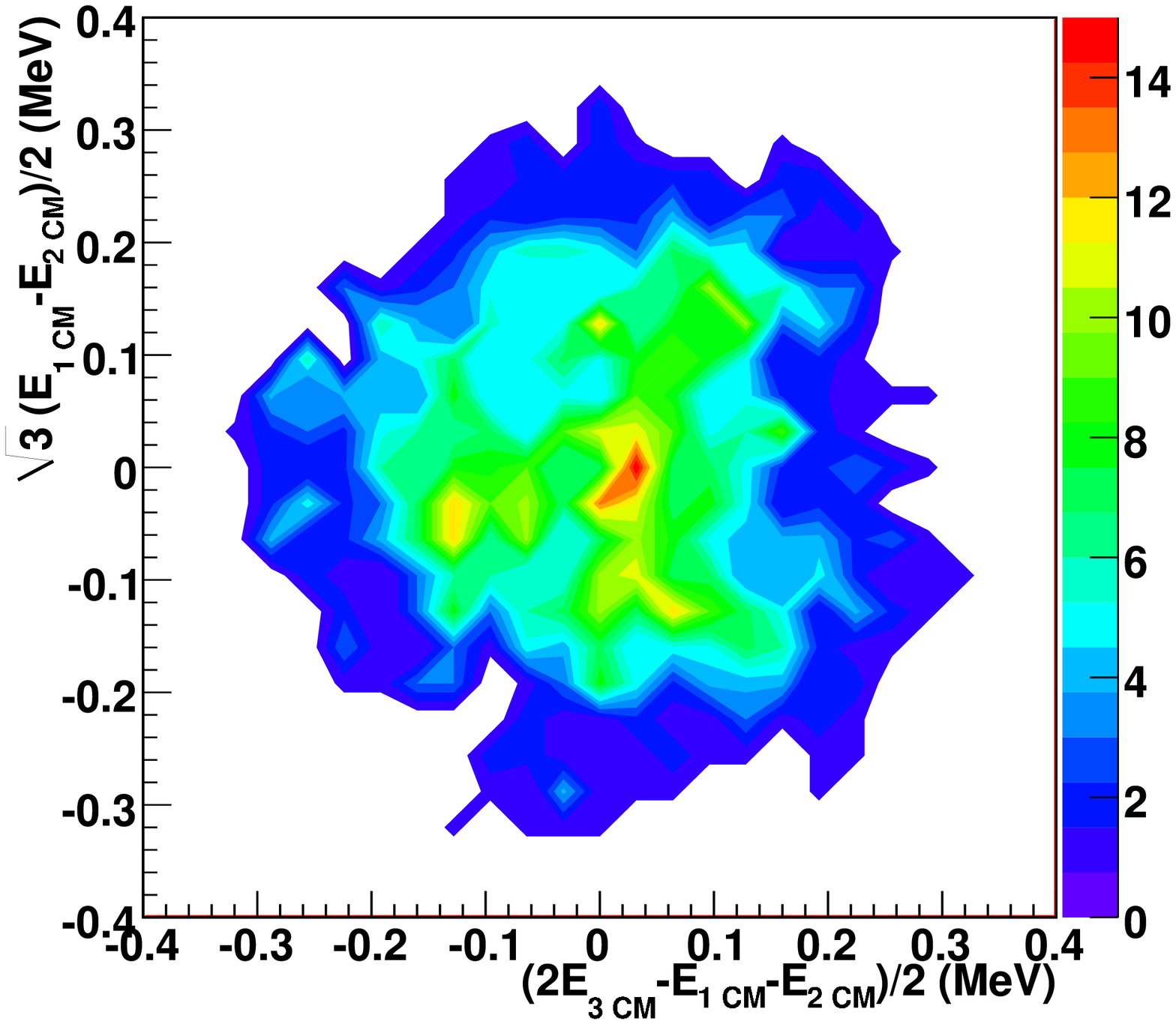,width=7.5cm,angle=90}
}
\caption{Three-$\alpha$ intra-event correlation function (left panel) 
and Dalitz plot (right panel) corresponding to QP events with
$m_{\alpha}=3$ and $7.375 \leq E_{ex} \leq 7.975$ MeV .
The uncorrelated yield is built such as to allow for decay through $^8$Be.
The dotted line marks the maximum RMS compatible with momentum
conservation.
}
\label{fig:Hoyle}
\end{figure}
The left panel of Fig. \ref{fig:Hoyle} presents the intra-event
correlation function corresponding to the QP events 
with $m_{\alpha}=3$ and 
$7.375 \leq E_{ex} \leq 7.975$ MeV, where
$m_{\alpha}$ is the alpha multiplicity 
and $E_{ex}$ denotes the excitation energy of the assumed decaying state.
As one may notice, around the energy of the Hoyle state above the $3\alpha$
threshold 
two peaks are observed. 
The first one, characterized by $\left<E_{\alpha} \right>=110$ keV
and $\sigma_{E_{\alpha}} \leq 25$ keV, corresponds, within our energy
uncertainty (calibration and direction of velocity vectors), to the equal
sharing of the available energy of the Hoyle state among the three
$\alpha$-particles. The second peak localized around
$\left<E_{\alpha} \right>=130$ keV
and $\sigma_{E_{\alpha}} \approx 90$ keV corresponds to the sharing 
of the available energy between the two $\alpha$s of $^8$Be and the remaining
$\alpha$ of 191 keV. Numerical simulations filtered by the software replica of
the detector confirm these preferential populations of the bi-dimensional
event representation. 
Even more, in the sequential decay case, they allow us to detect the 
relative angle between the velocity vectors of the $\alpha$s originating 
from $^8$Be and the one of $^8$Be. Thus, the extreme case in which the three
$\alpha$s are flying along the same direction produces a peak centered at
($\left< E_{\alpha}\right> \approx $130 keV, 
$\sigma_{E_{\alpha}}\approx $ 85 keV) while the region around
($\left< E_{\alpha}\right> \approx $90 keV, 
$\sigma_{E_{\alpha}} \approx$ 60 keV) is populated when the two $\alpha$s 
of $^8$Be are emitted on a direction perpendicular to the one of $^8$Be.
 
The competition between direct and sequential decay of the Hoyle state
previously reported by Freer \cite{freer} is testified also by the Dalitz plot
(Fig. \ref{fig:Hoyle}, right panel) which clearly manifests the typical
pattern of each decay mechanism.

Both methods allow for an estimation of the
relative probability of a sub-class of events with respect to any other. 
Let us focus on the particle condensation candidates. 
They correspond to the events localized at  
($\left<E_{\alpha} \right>=110$ keV,
$\sigma_{E_{\alpha}} \leq 25$ keV) whose number is 39.
Taken into account that the total number of events with
($m_{\alpha}=3$ and $7.375 \leq E_{ex} \leq 7.975$ MeV) 
is 1072 and the number of events with
($m_{\alpha}=1$, $m_{8Be}=1$ and $7.375 \leq E_{ex} \leq 7.975$ MeV ) is 900
it comes out that 2\% of the populated Hoyle states are compatible with
$\alpha$-particle condensation.

\begin{figure}[th]
\centerline{
\vspace*{-10pt}
\psfig{file=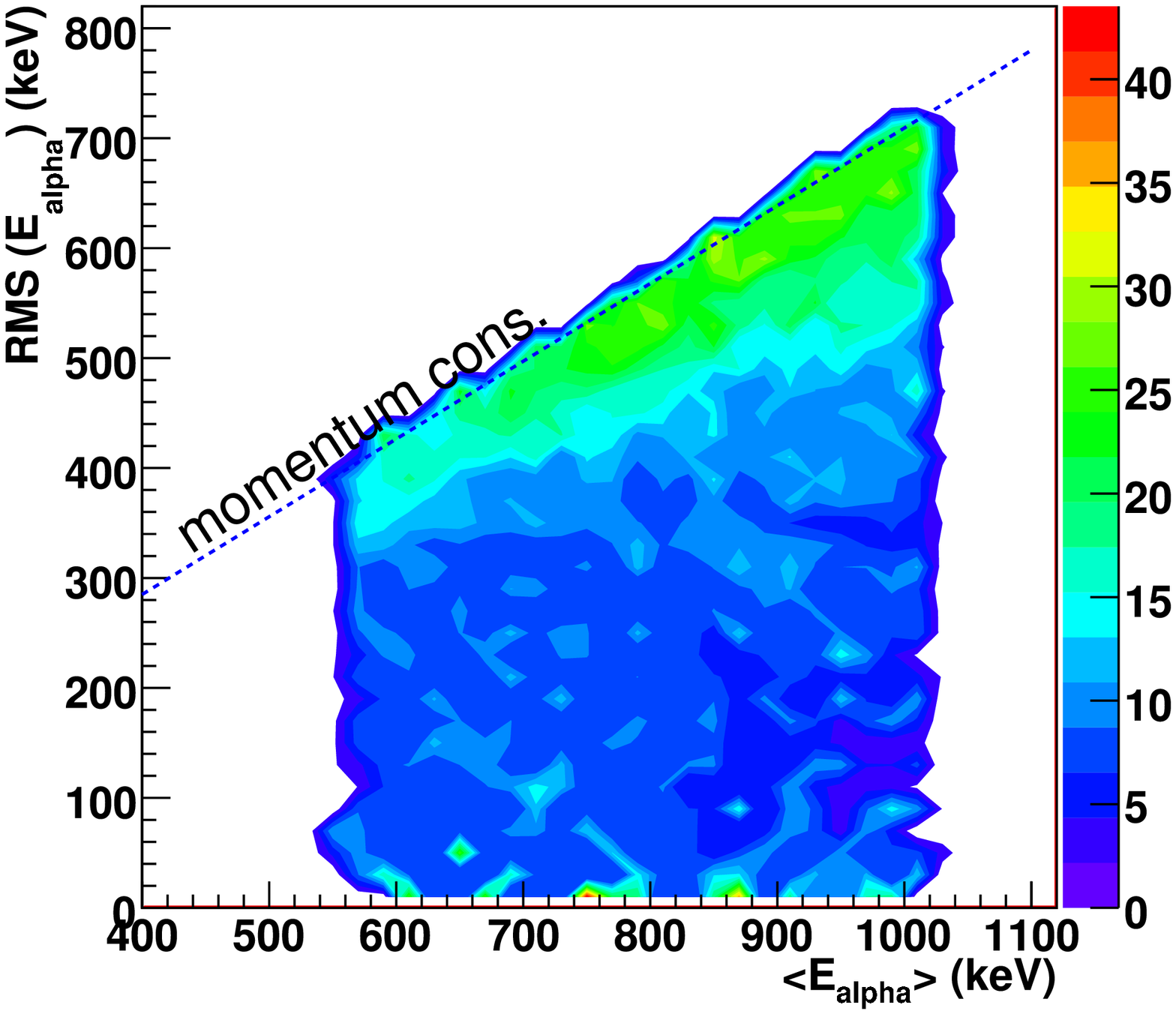,width=7.cm,angle=90}
\psfig{file=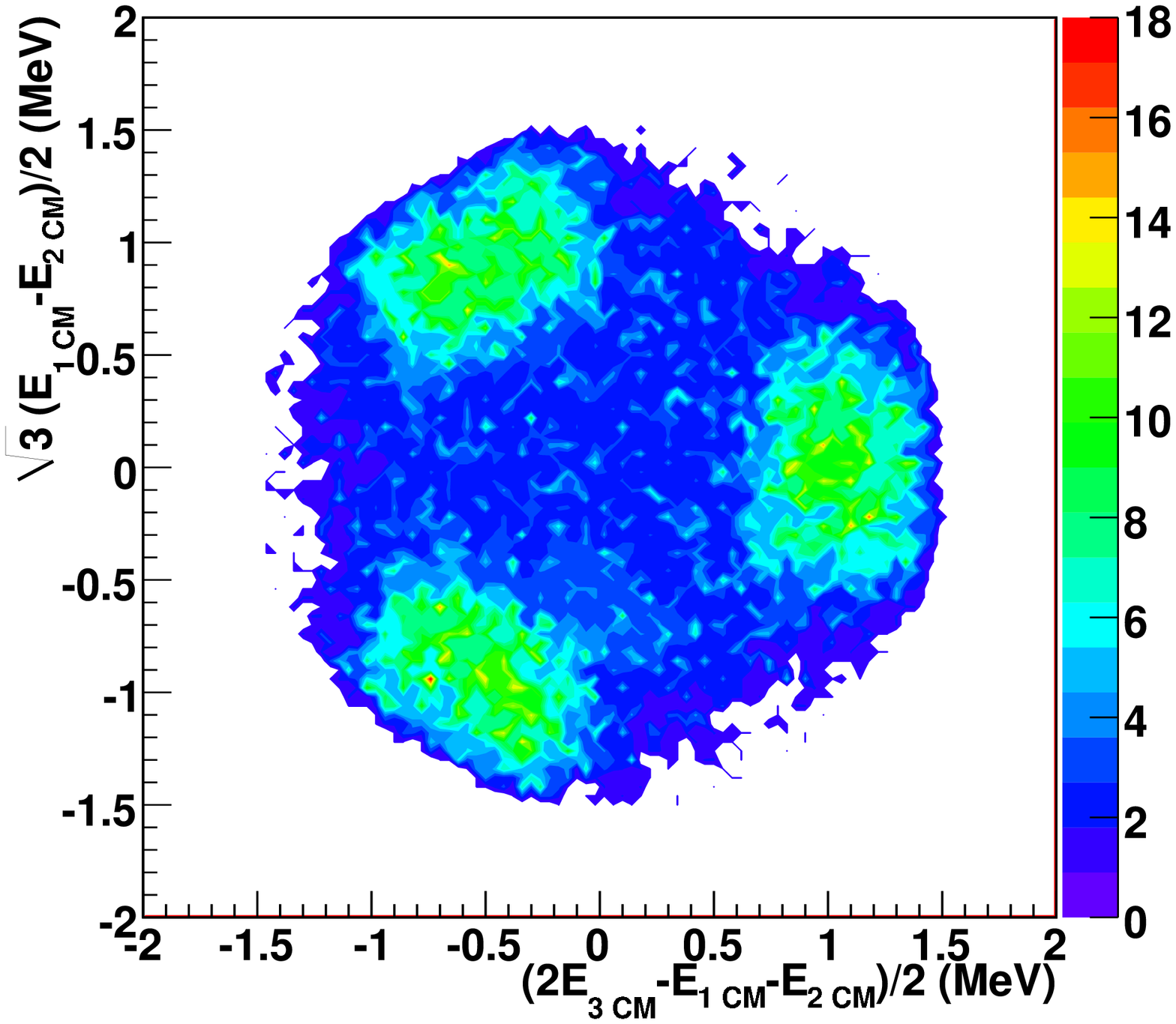,width=7.cm,angle=90}
}
\caption{Three-$\alpha$ intra-event correlation function (left panel) 
and Dalitz plot (right panel) corresponding to QP events with
$m_{\alpha}=3$ and $8.9 \leq E_{ex} \leq 10.3$ MeV.
The uncorrelated yield is built such to allow for decay through $^8$Be.
}
\label{fig:2nd}
\end{figure}
The same analyses may be performed in the case of the region around 
the $2_2^+$ state at 9.6 MeV of $^{12}$C. Fig.~\ref{fig:2nd} illustrates the intra-event
correlation function and Dalitz plot corresponding to the 
1.5$\cdot 10^4$ events
with $m_{\alpha}=3$ and $8.9 \leq E_{ex} \leq 10.3$ MeV.
The correlation function shows a broad peak which lies along the line
of maximum RMS compatible with momentum conservation. The Dalitz plot
manifests the symmetric three-bump structure typical for sequential decays.
Out of this, it comes out that no evidence is found in favor of condensation
compatible states in this case.

The intra-event correlation method, not restricted by the number of involved
particles, has been applied also to 4$\alpha$-events. Unfortunately, poor
statistics prevents us from drawing a definite conclusion. We nevertheless
mention that 4 states compatible with the above stated criteria of
condensation have been identified. A dedicated experiment is planned in the
near future.

To summarize, for the first time we have found experimental evidence in favor
of the $\alpha$-particle condensate nature of the Hoyle state of $^{12}$C. 

\section*{Acknowledgements}
One of the authors (Ad. R. R.) acknowledges the partial financial support 
from ANCS, Romania, under grant Idei nr. 267/2007.

\end{document}